\documentclass[epj]{svjour}
\usepackage{graphics}

\usepackage{amsmath}
\begin{document}
%\begin{fmffile}{diagrams}
\title{Shell model and deformed shell model spectroscopy of $^{62}$Ga}
\author{}
\author{P. C.~Srivastava$^1$ \thanks{e-mail: pcsrifph@iitr.ac.in}, R. Sahu$^2$ 
\and V.K.B. Kota$^3$}
\institute{Department of Physics, Indian Institute of Technology, Roorkee 247 667, 
India \and Physics Department, Berhampur University,
Berhampur - 760 007, Odisha, India \and Physical Research Laboratory, Ahmedabad 
380 009, India}
\date{\today}

%========================================================================
\abstract{   
In the present work we have reported comprehensive analysis of recently
available experimental data [H.M. David et al., Phys. Lett. B {\bf 726}, 665
(2013)] for high-spin states up to $17^+$ with $T=0$ in the odd-odd $N=Z$
nucleus $^{62}$Ga using shell model calculations within the full
$f_{5/2}pg_{9/2}$ model space and deformed shell model based on Hartee-Fock
intrinsic states in the same space.  The calculations have been performed using
jj44b effective interaction developed recently by B.A. Brown and A.F. Lisetskiy
for this  model space. The results obtained with the two models are  similar and
they are in reasonable agreement with experimental data. In addition to the $T=0$
and $T=1$ energy bands, band crossings and  electromagnetic transition
probabilities, we have also calculated the pairing energy in shell model and
all these compare well with the available theoretical results.}

\PACS{
      {21.60.Cs}{Shell model}, {21.10.Hw}  {Spin, parity, and isobaric spin} 
     }

%\pacs{21.60.Cs, 27.50.+e}  
\authorrunning{P.C. Srivastava}
\maketitle

\maketitle
%\nonindent
%%%%%%%%%%%%%%%%%%%%%%%%%%%%%%%%%%%%%%%%%%%%%%%%%%%%%%%%%%%
\section{Introduction}

In the recent past there are many experimental and theoretical investigations  
for heavy $N=Z$ nuclei. In the case of even-even $N=Z$ nuclei from $^{68}$Se
($Z=34$) to $^{88}$Ru ($Z=44$) many interesting phenomena have been observed.
For example, $^{68}$Se exhibits oblate shape in the ground state \cite{se68} and
in the case of $^{72}$Kr \cite{72kr} shape-coexistence has been observed. On the
other hand, the nuclei $^{76}$Sr\cite{76sr} and $^{80}$Zr\cite{80zr} have large
ground-state  deformation. Also, the even-even N=Z nuclei are waiting point
nuclei for $rp$-process nucleosynthesis. As we move further, there is a decrease
in deformation as seen for example in  $^{84}$Mo\cite{84mo} and
$^{88}$Ru\cite{88ru}. However, more interesting are the N=Z odd-odd nuclei as
they will allow us to investigate isospin effects and in particular about $T=0$ vs
$T=1$ pairing. As a consequence there are continuous experimental efforts to
study in detail odd-odd N=Z nuclei in the A $\sim$ 60-80 region
\cite{Vin,rudolph,david,as66,br70,rb74a,rb74b,rb74c,y78,62gaprl}.  In this paper we will
consider $^{62}$Ga where there are new and more complete data \cite{david}
obtained recently  by using heavy-ion fusion-evaporation reaction
$^{40}$Ca($^{24}$Mg,$pn$)$^{62}$Ga near the Coulomb barrier with ATLAS
accelerator and Gammasphere array. These data go much beyond the data reported
previously in 1998 \cite{Vin} and 2004 \cite{rudolph} on $T=0$ and $T=1$ levels
in this nucleus. Theoretical results for low-lying  T=0 and T=1 states, with low
spins, of $^{62}$Ga obtained using SM \cite{sm2}, DSM \cite{dsm} and IBM-4 
\cite{sm2} are available in the literature. In the previous
SM and DSM calculations
realistic G-matrix interaction with a phenomenologically adjusted monopole part
as given by the Madrid-Strasbourg group \cite{No-01} has been used.    
More recently,  Shell Model
using Madrid-Strasbourg effective interaction and Cranked Nilsson-Strutinsky
model calculations for yrast states of $^{62}$Ga have been performed 
\cite{npa}.  The aim
of the present study is to explain, more comprehensively, the recent
experimental  data for $^{62}$Ga using shell model (SM) calculations and
also extend the deformed shell model (DSM) (based on Hartree-Fock states)
calculations reported in the past for this nucleus \cite{dsm} both with a more
recently introduced effective interaction jj44b \cite{jj44b}.
In the present work we have not only focused on low-lying levels but also on
high-spin states up to $17^+$ with $T=0$. 

%This work will also add more
%information in the previous theoretical study on  $^{62}$Ga \cite{npa} where SM
%using Madrid-Strasbourg effective interaction and Cranked Nilsson-Strutinsky
%model results for yrast states are presented.

The paper is organized as follows. Section 2 gives some details regarding the
effective interaction used and about DSM. Section 3 gives results from SM and
DSM and their comparison with experimental data. Comparison with previous
theoretical studies is given in section 4. Finally, concluding remarks  are
drawn in section 5.

\section{Method of Calculations}

In the present work for both SM and DSM, $^{56}$Ni is taken as the inert core
with the spherical orbits $2p_{3/2}$, $1f_{5/2}$, $2p_{1/2}$ and $1g_{9/2}$ as
active orbits. The jj44b interaction due to Brown and Lisetskiy \cite{jj44b} was
employed in  the calculations. This interaction was developed by fitting 600
binding energies and excitation energies from nuclei with $Z = 28-30$ and $N =
48-50$. Here, 30 linear combinations of $JT$ coupled two-body matrix elements
(TBME) are varied giving the rms deviation of about 250 keV from  experiment.
The single particle energies (spe) are taken to be \cite{jj44b} -9.6566,
-9.2859, -8.2695 and -5.8944 MeV for the $p_{3/2}$, $f_{5/2}$, $p_{1/2}$ and
$g_{9/2}$ orbits respectively. The shell model calculations are carried
out using the shell model code NuShell \cite{nushell}. The maximum 
matrix dimension in $M$-scheme is for $0^+$ states and it is $91564$.

Turning to DSM, for a given nucleus, starting with a model space consisting of
the given set of single particle (sp) orbitals and effective two-body
Hamiltonian (TBME + spe), the lowest energy intrinsic states are obtained by
solving the Hartree-Fock (HF) single particle equation self-consistently.
Excited intrinsic configurations are obtained by making particle-hole
excitations over the lowest intrinsic state.  These intrinsic states 
$\chi_K(\eta)$ do not have definite angular momenta.  and states of good angular
momentum projected from an intrinsic state $\chi_K(\eta)$ can be written in the
form

\begin{equation}
\psi^J_{MK}(\eta) = \frac{2J+1}{8\pi^2\sqrt{N_{JK}}}\int d\Omega D^{J^*}_{MK}(\Omega)R(\Omega)| \chi_K(\eta) \rangle 
\label{eq.proj}
\end{equation}\\

where $N_{JK}$ is the normalization constant given by

\begin{equation}
N_{JK} = \frac{2J+1}{2} \int^\pi_0 d\beta \sin \beta d^J_{KK}(\beta)\langle \chi_K(\eta)|e^{-i\beta J_y}|\chi_K(\eta) \rangle 
\label{eqn2}
\end{equation}\\

In Eq. (\ref{eq.proj}) $\Omega$ represents the Euler angles ($\alpha$, $\beta$,
$\gamma$), $R(\Omega)$ which is equal to exp($-i\alpha J_z$)exp($-i\beta
J_y$)exp( $-i\gamma J_z$) represents the general rotation operator.  The good
angular momentum states projected from different intrinsic states are not in
general orthogonal to each other.  Hence they are orthonormalized and then band
mixing calculations are performed.  In addition, as required for N=Z nuclei,
isospin projection is also included  in DSM. For details see for example 
\cite{SSK,KS1,KS2,KS3}. DSM is well established to be a successful model for
transitional nuclei (with A=60-90) when sufficiently large number of intrinsic
states are included in the band mixing calculations. 

For $^{62}$Ga nucleus, fig. 1 gives the HF single particle (sp) spectrum (the
states are labeled by $|k_\alpha \rangle$ where the $\alpha$ label
distinguishes different states with the same $k$ value)  for both prolate and
oblate solution. As seen from the figure, the prolate solution is lowest.  The
lowest HF intrinsic state from the prolate solution consists of two protons  and
two neutrons occupying the lowest $k=1/2^-$ state and the last unpaired odd
proton and neutron occupying the next $k=1/2^-$ state.  The HF energy ($E$), the
mass quadrupole moment ($Q$) and band $K$-value are also shown in the figure.
For  the four nucleons (two protons and two neutrons) occupying the lowest 
$k=1/2^-$ sp state we have $T=0$.  Hence the isospin for $^{62}$Ga is determined
by the last proton and neutron. Thus the total isospin for the configuration
shown in fig. 1 is $T=0$ as the odd proton and odd neutron, for $K=1^+$, form a 
symmetric pair in the $k$-space (here and elsewhere in this paper symmetry in
$k$-space means symmetry in space-spin co-ordinates as $k$ contains both space
(orbital) and spin co-ordinates). Particle-hole excitations over the lowest HF
intrinsic state (from both prolate and oblate solutions) generate  excited HF
intrinsic states. There are 44 low-lying excited intrinsic states obtained by
particle-hole excitation up to 3 MeV excitation.  The HF intrinsic states are in
general admixtures of various isospin components. As mentioned above, the lowest
prolate and oblate HF intrinsic states will have $T = 0$.  If in an excited
intrinsic state, the unpaired proton occupies the single particle orbit
specified by the azimuthal quantum number $k_1$ and the unpaired neutron
occupies the state $k_2$, then one can also consider an intrinsic state where
the occupancies of the unpaired nucleons are reversed. By taking a linear
combination of these intrinsic states, one can construct intrinsic states which
are symmetric (or antisymmetric) in $k$-space co-ordinates and they correspond
to $T=0$ and $T=1$ states respectively. With this isospin projection, in the 
present calculation, there are 26 intrinsic states for $T=0$ and 18 for $T=1$
(total 44). All the configurations are listed in table 1. 
Then good angular momentum states are projected from all the $T
= 0$ intrinsic states and  a band mixing calculation is performed. Similar
procedure is also applied for the $T = 1$ intrinsic states.

\section{Results and discussions}

\subsection{Shell model results}

Fig. 2 shows comparison of recently available experimental data \cite{david}
with SM for the spectra with the lowest $T=1$ band and three higher $T=0$ bands
with maximum spin $17^+$. The agreements are reasonable. However, the SM gives
the excitation energy of the lowest $T=0$ level (with $1^+$) to be 148 keV
against the experimental value 571 keV. In order to bring out the
structure of these levels, in table 2 given are the dominant shell model
configuration (and its probability) in a given level and also the occupancies of
the four single particle orbits. In the lowest $T=0$ band, the high-spin levels
starting from $J^\pi=13^+$ have $g_{9/2}$ occupancy close to $2$ while the lower
levels are essentially from the $f_{5/2}p$ orbits. For example, the structure
changes from $1^+$ with dominant configuration $f_{5/2}^3p_{3/2}^3$  ($\sim$
14.66\%) to $17^+$ with dominant configuration $f_{5/2}^2p_{3/2}^2g_{9/2}^2$
($\sim$ 68.87\%).  This shows that as we move to higher $J$ values, the
$g_{9/2}^2$ orbital  start playing an important role in the structure. It is
also seen from table 2 that the second $T=0$ band is  essentially from
$f_{5/2}p$ orbits for spins up to $10^+$ while for the third  band $g_{9/2}$ is
important for the $10^+$ level. Turning to yrast T=1 states,  it is seen that
shell model gives very good agreement with available experimental data.  The
structure of $0^+$ to $10^+$ levels is mainly due to $(f_{5/2}p_{3/2}p_{1/2})^6$
configuration and the structure of low $J$ values is more fragmented in
comparison to high $J$ values. This is reflected from the change in the
probability $\sim$ 17.93\% ($0^+$) to $\sim$ 65.15\%  ($10^+$ ).

There is interest in the number of low-lying (say up to 2 or 3 MeV) states  in
odd-odd N=Z nuclei compared to the neighboring neutron-rich odd-odd nuclei. For
example, the $^{62}$Ga with N=Z has much lower number of levels up to 1.7 MeV
excitation compared to those in $^{64}$Ga ($\sim 30$ levels) and $^{68}$Ga ($\sim
60$ levels); see the discussion in \cite{david}. Because of this important
issue, we show in fig. 3 all the low-lying levels up to 3 MeV excitation.
It is seen that the sequence of lowest-lying  states is well reproduced
by shell model although the calculated level energies are compressed. 

Finally, some results for $B(E2)$'s are shown in table 3 and for $B(M1)$'s in 
table 4. For $T=0$ state the B(E2,$3_1^+ \rightarrow 1_1^+$) from shell model is
6.43 W.u. while the corresponding experimental value is 12 W.u. The  result may
improve if we slightly increase the effective charges. Similarly, the $B(M1)$ is
largest for the $0^+_1$ of $T=1$ to $1^+_1$ of $T=0$. Further discussion is
given below.

%%%%%%%%%%%%%%%%%%%%%%%%%%%%%%%%%%%%%%%%%%%%%%%%%%%%%%%
%\newpage
\begin{figure}
\resizebox{0.60\textwidth}{!}{
\includegraphics{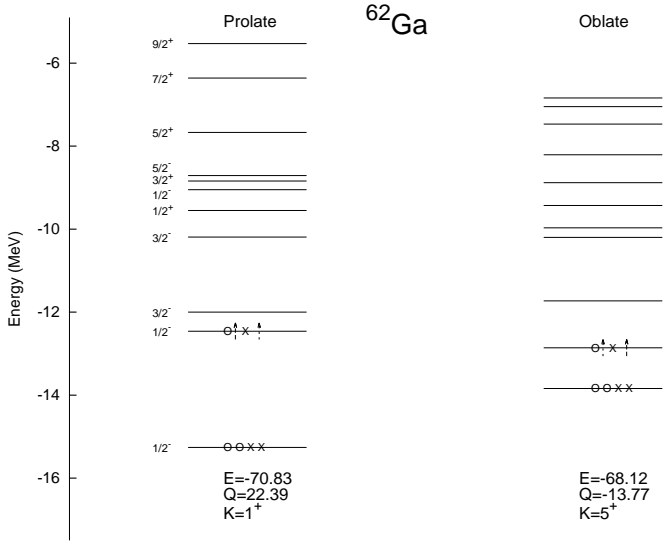} 
}
\caption{HF single particle spectra for $^{62}$Ga 
corresponding to lowest prolate and oblate configurations.  In the
figures circles represent protons and crosses represent neutrons. The
Hartree-Fock energy  ($E$) in MeV, mass quadrupole moment ($Q$) in units of
the square of the oscillator length parameter and the total $K$ quantum
number of the lowest intrinsic states are given in the figure.}
\label{dsm-hf}      
\end{figure}  
%%%%%%%%%%%%%%%%%%%%%%%%%%%%%%%%%%%%%%%%%%%%%%%%%%%%%%%%%%

\begin{table}
\begin{center}
{\small{
\caption{Intrinsic states used for $^{62}$Ga in DSM calculation.
For each  intrinsic state, given are the corresponding 
configurations, the total $K$-quantum number and isospin ($T$). 
The states 1 to 20 are prolate configurations and 21 to 26 oblate. 
Superscript $(2p,2n)$ implies that the orbit is occupied by two protons and two
neutrons and similarly the superscript $(p,n)$. In addition, the superscripts
$p(n)$ and $n(p)$ imply that the orbit(s) is(are) alternatively occupied by 
a proton and a neutron or a neutron and a proton.  }
\begin{tabular}{c|c|c|l}
\hline 
\hline
No. &  $\;\;K\;\;$ & $\;\;T$  & $\;\;\;$Configuration \\  
\hline
1.  &  $1^+$ &  0  & $(1/2^-)_1^{2p,2n}(1/2^-\uparrow)_2^{p,n}$  \\
2. &$0^+$&0,1& $(1/2^-)_1^{2p,2n}(1/2^-\uparrow)_2^{p(n)}(1/2^-\downarrow)_2^{n(p)}$\\
3.  &  $3^+$ &  0  & $(1/2^-)_1^{2p,2n}(3/2^-\uparrow)_1^{p,n}$  \\
4.  &  $0^+$ & 0,1 & $(1/2^-)_1^{2p,2n}(3/2^-\uparrow)_1^{p(n)}(3/2^-\downarrow)_1^{n(p)}$ \\
5.  &  $1^+$ & 0,1 & $(1/2^-)_1^{2p,2n}(3/2^-\uparrow)_1^{p(n)}(1/2^-\downarrow)_2^{n(p)}$ \\
6.  & $2^+$ & 0,1 & $(1/2^-)_1^{2p,2n}(3/2^-\uparrow)_1^{p(n)}(1/2^-\uparrow)_2^{n(p)}$  \\
7.  &  $3^+$ &  0  & $(1/2^-)_1^{2p,2n}(3/2^-\uparrow)_2^{p,n}$  \\
8.  & $0^+$ & 0,1 & $(1/2^-)_1^{2p,2n}(3/2^-\uparrow)_2^{p(n)}(3/2^-\downarrow)_2^{n(p)}$  \\
9.  & $1^+$ & 0,1 & $(1/2^-)_1^{2p,2n}(3/2^-\uparrow)_2^{p(n)}(1/2^-\downarrow)_2^{n(p)}$ \\
10.& $2^+$ & 0,1 & $(1/2^-)_1^{2p,2n}(3/2^-\uparrow)_2^{p(n)}(1/2^-\uparrow)_2^{n(p)}$ \\
11.& $0^+$ & 0,1 & $(1/2^-)_1^{2p,2n}(3/2^-\uparrow)_2^{p(n)}(3/2^-\downarrow)_1^{n(p)}$ \\
12.& $3^+$ & 0,1 & $(1/2^-)_1^{2p,2n}(3/2^-\uparrow)_2^{p(n)}(3/2^-\uparrow)_1^{n(p)}$ \\
13.  &  $1^+$ &  0  & $(1/2^-)_1^{2p,2n}(1/2^+\uparrow)_1^{p,n}$  \\
14.  &$0^+$&0,1& $(1/2^-)_1^{2p,2n}(1/2^+\uparrow)_1^{p(n)}(1/2^+\downarrow)_1^{n(p)}$\\
15.  &  $3^+$ &  0  & $(1/2^-)_1^{2p,2n}(3/2^+\uparrow)_1^{p,n}$  \\
16.  &$0^+$&0,1& $(1/2^-)_1^{2p,2n}(3/2^+\uparrow)_1^{p(n)}(3/2^+\downarrow)_1^{n(p)}$\\
17. &$1^+$&0,1& $(1/2^-)_1^{2p,2n}(3/2^+\uparrow)_1^{p(n)}(1/2^+\downarrow)_1^{n(p)}$ \\
18. &$2^+$&0,1& $(1/2^-)_1^{2p,2n}(3/2^+\uparrow)_1^{p(n)}(1/2^+\uparrow)_1^{n(p)}$\\
19.  &  $1^+$ &  0  & $(1/2^-)_2^{2p,2n}(1/2^-\uparrow)_1^{p,n}$  \\
20. &$0^+$&0,1& $(1/2^-)_2^{2p,2n}(1/2^-\uparrow)_1^{p(n)}(1/2^-\downarrow)_1^{n(p)}$\\
21.  &  $5^+$ &  0  & $(3/2^-)_1^{2p,2n}(5/2^-\uparrow)_1^{p,n}$  \\
22. &$0^+$&0,1& $(3/2^-)_2^{2p,2n}(5/2^-\uparrow)_1^{p(n)}(5/2^-\downarrow)_1^{n(p)}$\\
23.  &  $1^+$ &  0  & $(3/2^-)_1^{2p,2n}(1/2^-\uparrow)_1^{p,n}$  \\
24. &$0^+$&0,1& $(3/2^-)_1^{2p,2n}(1/2^-\uparrow)_1^{p(n)}(1/2^-\downarrow)_1^{n(p)}$\\
25.  &$3^+$&0,1& $(3/2^-)_1^{2p,2n}(5/2^-\uparrow)_1^{p(n)}(1/2^-\uparrow)_1^{n(p)}$  \\
26. &$2^+$&0,1& $(3/2^-)_1^{2p,2n}(5/2^-\uparrow)_1^{p(n)}(1/2^-\downarrow)_1^{n(p)}$\\
\hline\hline\end{tabular}
}}
\end{center}
\end{table}

\subsection{Deformed shell model results}

Figs. 2 and 3 shows comparison of DSM results with experimental data 
\cite{david} and SM for high spin states and  all low-lying levels respectively.
The agreements between experiment and DSM and also between DSM and SM are
reasonable. As discussed in \cite{KS1}, DSM can produce only the relative
energies in the $T=0$ levels and similarly for the $T=1$ levels. Following this,
all the $T=0$ levels are pushed up by 600 keV (experimental value is 571 keV
with respect to the lowest  $T=1$ level just as in \cite{KS1,KS2}).  For the
lowest $T=0$ band (see fig. 1) it is seen that there is a  band crossing at
$11^+$ with clear structural change from $13^+$.  The $13^+$, $15^+$ and $17^+$
mainly originate from the rotational aligned $K=1^+$ band obtained by placing a
proton (neutron) in a  $k=3/2^+$ and neutron (proton) in a $k=1/2^+$ orbit. This
corresponds to the configuration number 17 in table 1. This configuration given
by DSM is consistent with SM result as given table 2, i.e. the high-spin levels
from $13^+$ have $g_{9/2}$ occupancy $\sim 2$. The levels below $11^+$ are 
essentially from  $f_{5/2}p$ orbits (as in SM, see table 2)  dominated by the
deformed configurations \#1, \#2 and \#4 in table 1. As seen from the fig. 2,
there are two close-lying $9^+$ levels with $T=0$ (this is seen for $9^+$ in
experimental data) and the $9^+_1$ arise from the mixing of the configurations
\#1 and \#10 (close in structure to $7^+_1$) while $9^+_2$ arise from mixing of
configurations \#1 and \#6. The SM has also two $9^+$ levels as shown in fig.
2.  From the $B(E2)$ values shown in table 3 for $9^+_i \rightarrow 7^+_1$ it is
clear, as we expect large $B(E2)$ value with the levels belonging to the same
band, that the $9^+_1$ (5075 keV) of DSM should correspond to $9^+_2$ (5405 keV)
of SM. For the two $11^+$ levels in DSM, there is considerable mixing of various
configurations.  From the $13^+_1 \rightarrow 11^+_i$ $B(E2)$ values shown in
table 3 and that there is clear structural change after $11^+$ allow as to
conclude that the $11^+_2$ (6311 keV) of DSM corresponds to $11^+_2$ (7421 keV)
of SM (see fig. 2). With these correspondence between DSM and SM we expect the
$B(E2)$ for $11^+_2 \rightarrow 9^+_1$ of DSM should be close to that of $11^+_2
\rightarrow 9^+_2$ of SM. This is indeed seen in table 3 (DSM gives 0.4 W.u.
and SM gives 1.05 W.u.). Thus we have a good correspondence between the yrast
$T=0$ levels in DSM and SM.  

Going to the non-yrast $T=0$ bands shown in fig. 2, the first excited $2^+$ band
like structure is mainly from configurations \#5 and \#6  and the second $2^+$
band arises mainly from configuration \#1 shown in table 1. Going to the $T=1$
band, this arises mainly from configurations \#2 and \#4. It is also seen that
with increase in spin in the band, there more mixing of other deformed
configurations (therefore SM configurations at higher spins are more pure as
stated in Section 3.1).  The $B(E2)'s$ values from DSM  and SM are similar 
for the $T=1$ band and the collectivity starts decreasing from $6^+$. For the
$10^+ \rightarrow 8^+$, the DSM value is much smaller than SM value. In 
addition, table 4 gives $B(M1)$ values for some of yrast $T=1$ to yrast $T=0$
transitions. It is seen that the $0^+_1 \rightarrow 1^+_1$ transition is strong
and the DSM value is close to SM value.  Other transition strengths are much
smaller in both models. 

Turning to the low-lying levels, all levels (with $T=0$ and $T=1$) predicted by
DSM below 3 MeV excitation are compared with SM and experimental data in fig. 3.
The number of levels in the experiment, SM and DSM with $T=0$ up to $1.7$ MeV
are 7, 17 and 10. 
As mentioned before, the experimentally
observed level density up to $1.7$ MeV excitation in the neighboring odd-odd Ga
isotopes is much larger. Another important feature is that in the $T=0$ levels, 
the experimental data show a well defined gap of $\sim 600$ keV   above 1.575
MeV level. A similar gap is seen in both DSM and SM. Also, the $2_2^+$, $0^+_2$
and $3^+_1$ of $T=1$ shown in fig. 3 for DSM (also SM) are indeed seen in the
isobaric analogue nucleus $^{62}$Zn. 
\begin{table}
\begin{center}
\caption{Shell model configurations and occupancies for $T=0$ and $T=1$ levels.}
\vspace{0.2cm}
\label{tab:table1}
\resizebox{8.9cm}{!}{
\begin{tabular}{ c | c |c | c  } \hline %\hline
\hline 
T=0   & \% probability  & configurations  &   nucleon occupation numbers      \\ 
       &                 &                 &   $n_{lj}^\pi$ =  $n_{lj}^\nu$ ($f_{5/2}$,$p_{3/2}$, $p_{1/2}$,$g_{9/2}$)   \\
\hline
$1^+$  & 14.66        &     $f_{5/2}^3$ $p_{3/2}^3$ $p_{1/2}^0$ $g_{9/2}^0$         &    1.71~ 3.08~ 1.00~ 0.21      \\
$3^+$  & 24.18         &     $f_{5/2}^2$ $p_{3/2}^3$ $p_{1/2}^1$ $g_{9/2}^0$          &    2.12~ 2.76~ 0.92~ 0.21      \\
$5^+$  & 30.30        &     $f_{5/2}^2$ $p_{3/2}^4$ $p_{1/2}^0$ $g_{9/2}^0$         &    2.22~ 2.74~ 0.79~ 0.24      \\
$7^+$  & 32.38        &     $f_{5/2}^2$ $p_{3/2}^3$ $p_{1/2}^1$ $g_{9/2}^0$        &    2.21~ 2.82~ 0.78~ 0.19      \\
$9^+$  & 32.94       &     $f_{5/2}^2$ $p_{3/2}^4$ $p_{1/2}^0$ $g_{9/2}^0$         &    2.32~ 2.97~ 0.59~ 0.11      \\
$13^+$  & 34.78         &    $f_{5/2}^3$ $p_{3/2}^1$ $p_{1/2}^0$ $g_{9/2}^2$          &    2.29~ 1.16~ 0.49~ 2.05      \\
$11^+$  & 94.59        &     $f_{5/2}^4$ $p_{3/2}^2$ $p_{1/2}^0$ $g_{9/2}^0$          &    3.94~ 1.94~ 0.01~ 0.11      \\
$15^+$  & 36.50        &     $f_{5/2}^2$ $p_{3/2}^1$ $p_{1/2}^1$ $g_{9/2}^2$         &    2.22~ 1.25~ 0.50~ 2.03      \\
$17^+$  & 68.87         &     $f_{5/2}^2$ $p_{3/2}^2$ $p_{1/2}^0$ $g_{9/2}^2$          &    2.33~ 1.65~ 0.00~ 2.02     \\
\hline 
$2^+$  & 14.57         &     $f_{5/2}^1$ $p_{3/2}^4$ $p_{1/2}^1$ $g_{9/2}^0$         &    1.87~ 2.97~ 0.94~ 0.21      \\
$4^+$  & 15.27         &     $f_{5/2}^2$ $p_{3/2}^3$ $p_{1/2}^1$ $g_{9/2}^0$         &    2.02~ 2.91~ 0.84~ 0.23      \\
$6^+$  & 17.89        &    $f_{5/2}^3$ $p_{3/2}^3$ $p_{1/2}^0$ $g_{9/2}^0$         &    2.62~ 2.43~ 0.70~ 0.25     \\
$8^+$  & 45.66       &     $f_{5/2}^3$ $p_{3/2}^3$ $p_{1/2}^0$ $g_{9/2}^0$         &    3.04~ 2.27~ 0.45~ 0.23     \\
$10^+$  & 69.08         &     $f_{5/2}^3$ $p_{3/2}^3$ $p_{1/2}^0$ $g_{9/2}^0$         &    3.09~ 2.64~ 0.16~ 0.10      \\
\hline 
$2^+$  & 16.24         &     $f_{5/2}^3$ $p_{3/2}^2$ $p_{1/2}^1$ $g_{9/2}^0$         &    2.44~ 2.45~ 0.89~ 0.22     \\
$4^+$  & 18.51         &     $f_{5/2}^0$ $p_{3/2}^5$ $p_{1/2}^1$ $g_{9/2}^0$          &    1.66~ 3.19~ 0.92~ 0.22      \\
$6^+$  & 30.12        &     $f_{5/2}^3$ $p_{3/2}^3$ $p_{1/2}^0$ $g_{9/2}^0$         &    2.29~ 2.93~ 0.58~ 0.20      \\
$8^+$  & 26.35        &     $f_{5/2}^2$ $p_{3/2}^3$ $p_{1/2}^1$ $g_{9/2}^0$         &    3.03~ 2.42~ 0.44~ 0.11      \\
$10^+$  & 31.27         &     $f_{5/2}^4$ $p_{3/2}^0$ $p_{1/2}^0$ $g_{9/2}^2$       &    2.90~ 0.91~ 0.28~ 1.90      \\
\hline
\hline
T=1   &   &        &          \\
\hline
$0^+$  & 17.93         &    $f_{5/2}^2$ $p_{3/2}^4$ $p_{1/2}^0$ $g_{9/2}^0$         &    1.89~ 2.85~ 0.81~ 0.43      \\
$2^+$  & 14.07         &     $f_{5/2}^2$ $p_{3/2}^3$ $p_{1/2}^1$ $g_{9/2}^0$         &    2.09~ 2.65~ 0.90~ 0.35      \\
$4^+$  & 17.23         &     $f_{5/2}^3$ $p_{3/2}^3$ $p_{1/2}^0$ $g_{9/2}^0$         &    2.53~ 2.41~ 0.74~ 0.33      \\
$6^+$  & 17.58        &     $f_{5/2}^3$ $p_{3/2}^3$ $p_{1/2}^0$ $g_{9/2}^0$          &    2.78~ 2.35~ 0.64~ 0.24      \\
$8^+$  & 24.55        &     $f_{5/2}^3$ $p_{3/2}^3$ $p_{1/2}^0$ $g_{9/2}^0$         &    2.86~ 2.50~ 0.44~ 0.19      \\
$10^+$  & 65.15         &    $f_{5/2}^3$ $p_{3/2}^3$ $p_{1/2}^0$ $g_{9/2}^0$          &    3.28~ 2.59~ 0.14~ 0.11      \\
\hline
\hline
\end{tabular}}
\end{center}
\end{table}

\begin{table}
\begin{center}
\caption{$B(E2)$ values (in W.u.)  in SM and DSM obtained using
effective charges $e_p=1.5e$ and $e_n=0.5e$. Experimental value (shown in the
last column) is taken from the NNDC database.}
\vspace{0.2cm}
\label{tab:table2}
\resizebox{!}{!}{
\begin{tabular}{ c | c |c | c  } \hline\hline
$I_f^+ \rightarrow I_i^+$ & DSM   & SM  & EXPT.\\ \hline
T=0 (Isoscalar) &   &              &         \\\hline
$3_1^+ \rightarrow 1_1^+$  & 24.7  & 6.46             & 12$^{+6}_{-3}$\\ 

$5_1^+ \rightarrow 3_1^+$  & 30.9   &     14.43         &         \\

$7_1^+ \rightarrow 5_1^+$  & 29.2   &  13.92           &         \\ 

$9_1^+ \rightarrow 7_1^+$  & 17.5  &    0.0029          &         \\ 

$9_2^+ \rightarrow 7_1^+$  & 0.02   &   7.77           &         \\ 

$11_1^+ \rightarrow 9_1^+$  &0.08 &     16.17         &         \\ 

$11_2^+ \rightarrow 9_1^+$  & 0.4  &     0         &         \\

$11_1^+ \rightarrow 9_2^+$  & 1.1   &   0          &         \\ 

$11_2^+ \rightarrow 9_2^+$  & 5.9  &     1.05         &         \\ 

$13_1^+ \rightarrow 11_1^+$  & 0.005   &    16.21        &         \\

$13_1^+ \rightarrow 11_2^+$  & 0.006   &   0           &         \\

$15_1^+ \rightarrow 13_1^+$  & 36.0  &   12.49          &         \\

$17_1^+ \rightarrow 15_1^+$  & 20.5  &    8.49          &         \\
\hline
T=1  (Isoscalar)&   &              &         \\
\hline
$2_1^+ \rightarrow 0_1^+$  & 51.6   &    42.57        &         \\

$4_1^+ \rightarrow 2_1^+$  & 64.7   &     53.91        &         \\

$6_1^+ \rightarrow 4_1^+$  & 45.9   &    52.52         &         \\

$8_1^+ \rightarrow 6_1^+$  & 20.6  &   23.13          &         \\

$10_1^+ \rightarrow 8_1^+$  & 6.5  &     18.12        &         \\
\hline
\hline
\end{tabular}}
\end{center}
\end{table}
\begin{table}[h]
\begin{center}
\caption{$B(M1)$ values in $\mu_N^2$. Here $g_s=g_{free}$ used in both SM and
DSM calculations.}
%\vspace{0.2cm}
\label{tab:table3}
%\resizebox{7.8cm}{4.5cm}{
\begin{tabular}{ c | c |c  } 
\hline \hline

$I_f^+ \rightarrow I_i^+$ & DSM   & SM \\ \hline

T=1 $\rightarrow$ T=0 (Isovector)&   &              \\
\hline
$0_1^+ \rightarrow 1_1^+$  & 1.2  &  1.39         \\

$2_1^+ \rightarrow 1_1^+$  & 0.005 & 0.14        \\

$2_1^+ \rightarrow 3_1^+$  & 0.13   & 0.01        \\

$4_1^+ \rightarrow 3_1^+$  & 0.08  & 0         \\

$4_1^+ \rightarrow 5_1^+$  & 0.13  & 0              \\

$6_1^+ \rightarrow 7_1^+$  & 0.06  & 0.013         \\

$8_1^+ \rightarrow 7_1^+$  & 0.41  & 0.16          \\
\hline
\hline

\end{tabular}
\end{center}
\end{table}

%\newpage
\begin{figure*}
\resizebox{0.95\textwidth}{!}{
\includegraphics{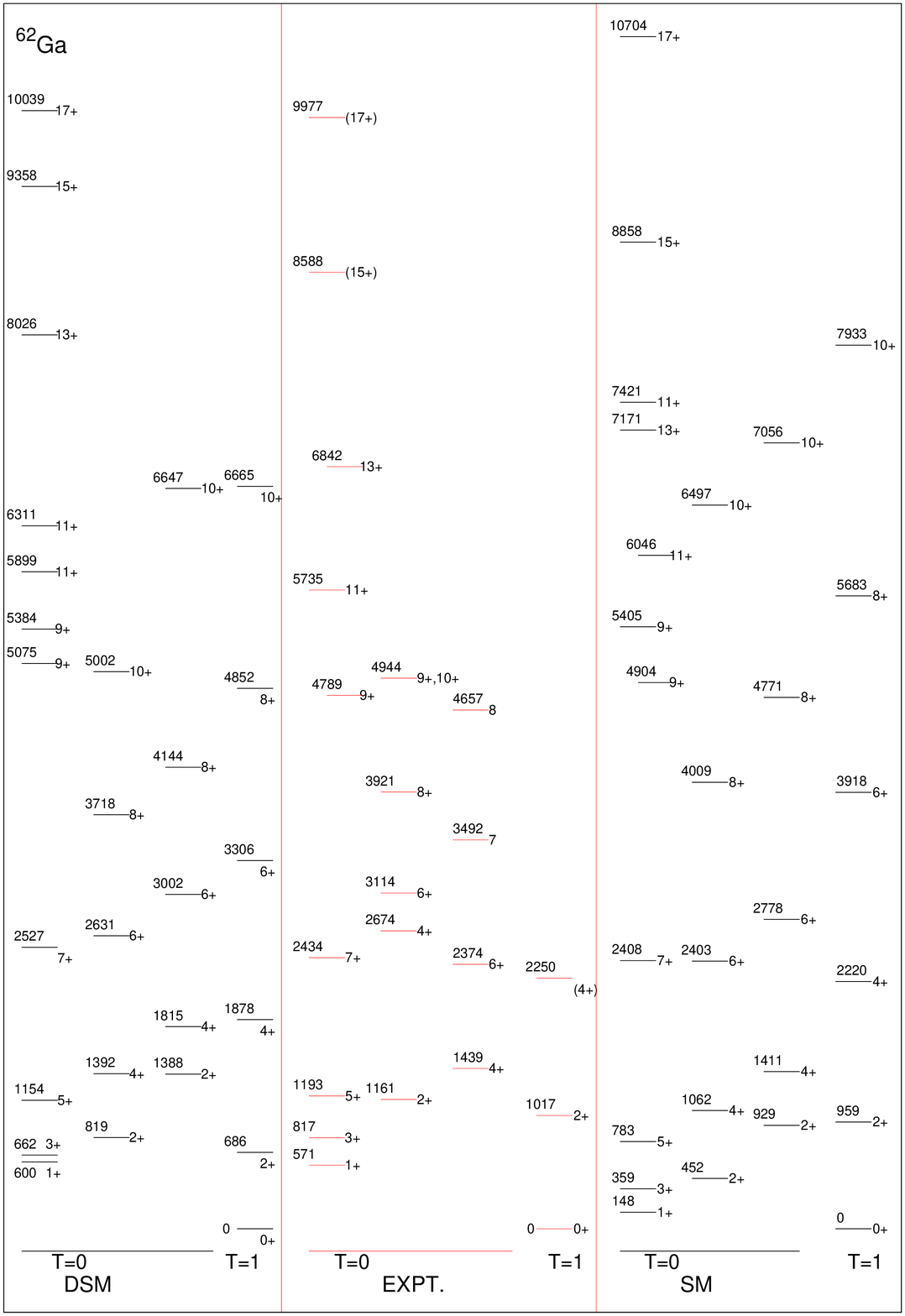} 
}
\caption{Comparison of  deformed shell model and shell-model results using jj44b 
interaction with experimental data. Energies are in keV.}
\label{calculs_smdsm}      
\end{figure*}
%%%%%%%%%%%%%%%%%%%%%%%%%%%%%%%%%%%%%%%%%%%%%%%%%%%%%%%%%%
%\newpage
\begin{figure*}
\resizebox{1.0\textwidth}{!}{
\includegraphics{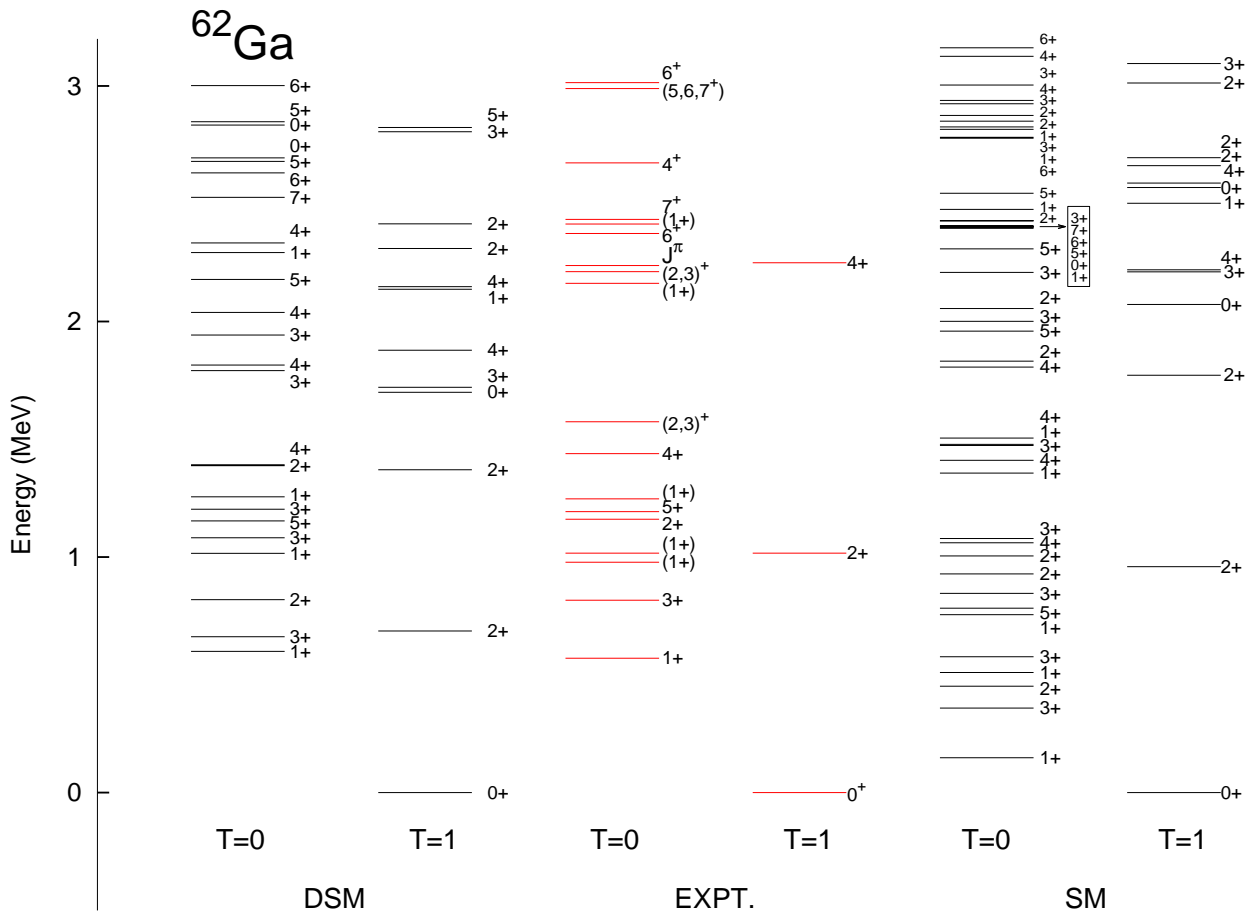} 
}
\caption{Comparison of experimental data \cite{david,62gaprl} with deformed shell model and 
shell-model results obtained using jj44b interaction for low-lying levels up to
$3$ MeV excitation. In the figure, the  results for SM and DSM appear to be different and
this is mainly because the energy  of the lowest $T=0$ state is different in the
two models. If they are aligned with the experimental level, it is seen that up
to about 2 MeV excitation the two models are close to experiment.
%For SM we have only reported maximum three eigen values for
%a given $J$.
}
\label{calculs_low}      
\end{figure*}

%\newpage
\begin{figure*}[bt]
\resizebox{1.0\textwidth}{!}{
\includegraphics{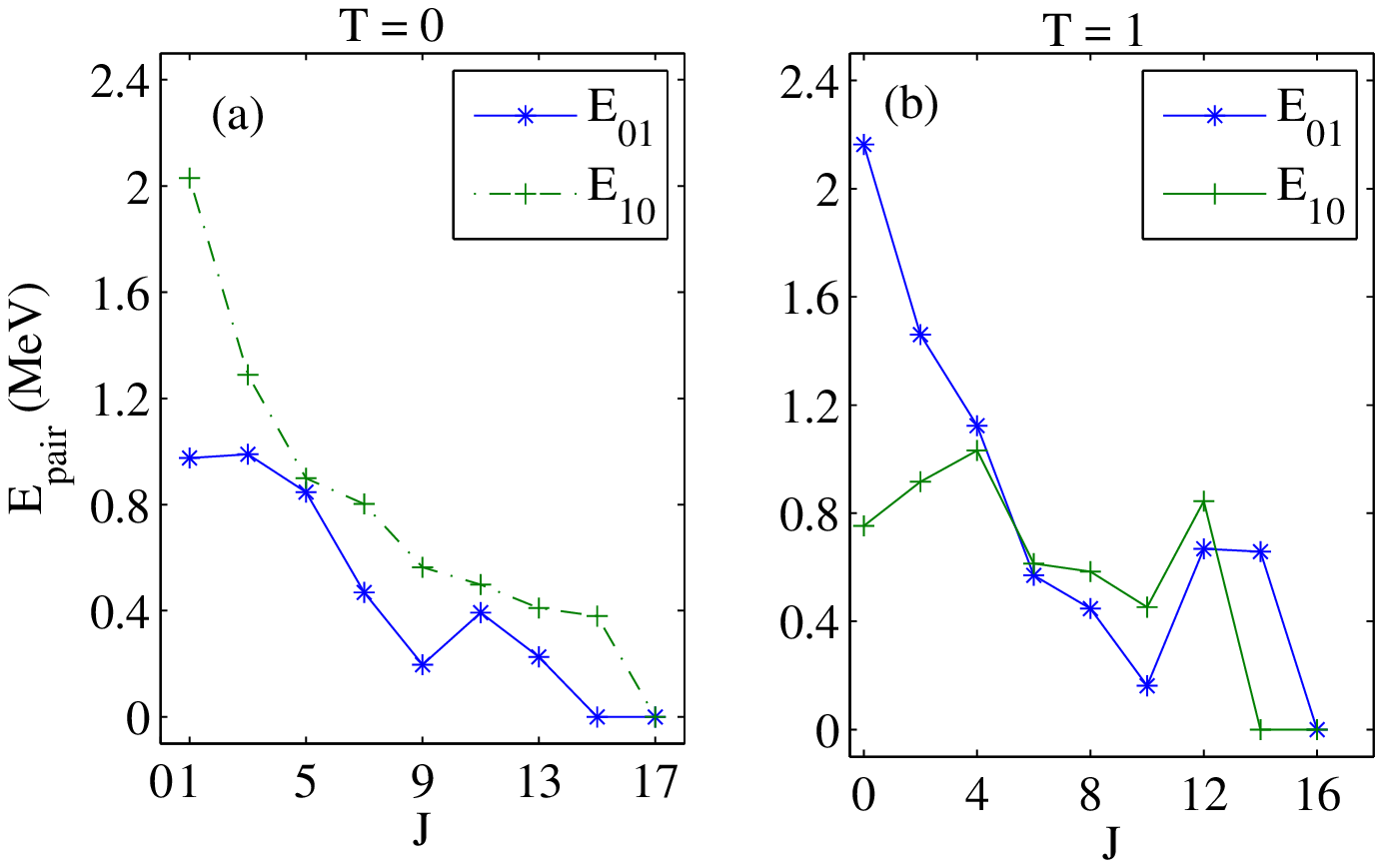} 
}
\caption{Pairing energies from the SM calculation for $T =0$ odd spin states and $T = 1$ even spin states.
Here $E_{01}$ and $E_{10}$ are isovector and isoscalar pairing, respectively.
}
\label{pairing}      
\end{figure*}

%\newpage
\section{Comparison with previous theoretical studies}

\subsection{Spectroscopy}

As discussed briefly in the introduction, previous theoretical studies on
$^{62}$Ga using shell model, IBM-4 and   Cranked Nilsson-Strutinsky model (CNS)
were reported in refs. \cite{sm2,npa}. In \cite{sm2} through a mapping, boson
Hamiltonian was derived microscopically from a realistic shell model
interaction. For low-lying levels one-to-one correspondence between shell model
and  IBM-4 was obtained. With IBM-4 the first excited 1$^+$ and 3$^+$ levels
were correctly reproduced. Also, IBM-4 with MS interaction (ref. \cite{npa} )
reproduces correctly  the $T=0$ and $T=1$ band head separation.  In present
study,  jj44b interaction predicts this separation to be 148 keV compared to the
experimental value 571 keV. However the relative spacings are quite well
reproduced in our study  as in \cite{npa}.  In the previous shell model study
\cite{sm2} up to 2.5 MeV only four 1$^+$ ($T=0$) levels were reported while in
the recent experimental work \cite{62gaprl} six plausible 1$^+$ levels were
reported. However, our shell model calculations with jj44b interaction generates
six 1$^+$ ($T=0$) levels as seen in experiment.  The first 2$^+$ ($T=1$) level
is  compressed in \cite{sm2} while our shell model results are much better with
difference of only 58 keV. 

The change in occupancy of the levels as we go to higher spins  provide
information regarding the change in structure of these levels. As seen in table
2, we find that all the $T=0$ odd spin states up to $J=11^+$ are  predominantly
from $f_{5/2}$ orbital. The occupation of $f_{5/2}$ orbital also increases with
spin.  However, for higher spin states the $g_{9/2}$ occupancy is about 2. As
discussed before, the DSM also predicts a band crossing  where a $K=1^+$  band 
obtained by excitation of two particles to $g_{9/2}$ crosses the  ground band at
$J=13^+$ and becomes yrast. Band crossing is the mechanism for back bending and
both our SM and DSM reproduces this phenomenon. Similar results were also
obtained by Juodagalvis and \AA berg (ref. \cite{npa}) in their shell model and
CNS calculations. However the transition point  was at spin $J=9^+$ in their
calculation. As seen in table 2, the occupancy of the $T=1$ even spin states 
are predominately from $f_{5/2}$ with occupancy of $f_{5/2}$ increasing with
spin up to $J=10^+$. Similar trend was seen in ref. \cite{npa}. The importance
of inclusion of $f_{7/2}$ orbital in the shell model space is reported in ref.
\cite{npa} from  the result of occupancy of the orbitals in  Cranked
Nilsson-Strutinsky model (CNS).  They argued that the smaller value of
$|Q_{spec}|$ in SM result (fig. 9, ref. \cite{npa}) reflects the importance of
inclusion of $f_{7/2}$ orbital. 

As indicated in fig. 8 ( ref. \cite{npa} ), our $B(E2)$'s value are also larger
for $T=1$ band compared to $T=0$ band.  The shell model in ref. \cite{npa}
predicts $B(E2,3^+ \rightarrow 1^+)$  to be about 100 $e^2fm^4$. This value in
the present shell model is 94. The trend for higher spins is also reasonably
well reproduced. The CNS model predicts faster decrease in $B(E2)$ as one goes
to higher spins.  For example $B(E2, 15_1^+ \rightarrow 13_1^+$) and $B(E2,
17_1^+ \rightarrow 15_1^+)$ in our calculation  are 182 and  124 $e^2fm^4$.
These values are larger than the values given in ref. \cite{npa}. The $B(E2)$
values for $T=1$ even spin states show a slow rising trend and  finally
decreases for $10^+$ as in ref. \cite{npa}. 

In ref. \cite{npa} it was concluded that for the low spin parts of the bands
with $T=0$ and $T=1$ corresponds to triaxially deformed states with the rotation
taking place around the shortest and intermediate axis, respectively. Our result
also supporting this conclusion because of different values of electromagnetic
properties of these two bands (see, tables 3 and 4) and also because DSM
description shows mixing of several intrinsic states.

\subsection{Pairing Energy}

We have also calculated the pairing energy following the procedure laid down
in ref. \cite{povesplb}. The two body matrix elements for the isovector pairing
$P01$ and isoscalar pairing $P10$ in jj coupling scheme are,
\begin{equation}
\begin{array}{l}
\lefteqn{\left\langle j_a j_b JT|P01|j_c j_d JT \right\rangle  } \\
\\
=G \sqrt{(j_a+1/2)(j_c + 1/2)} \delta_{ab} \delta_{cd} \delta_{J0}
\delta_{T1}\;,
\end{array}
\end{equation}
\begin{equation}
\begin{array}{l}
\lefteqn{\left\langle j_a j_b JT|P10|j_c j_d JT \right\rangle } \\
\\
=G\frac{2(-1)^{j_a-j_c}}{\sqrt{1+\delta_{ab}}\sqrt{1+\delta_{cd}}} \\
\\
\times \sqrt{(2j_a+1)(2j_b+1)(2j_c+1)(2j_d+1)} \\
\\
\times \left\{ \begin{array}{ccc}
1/2 & j_a & l_a\\
j_b & 1/2 & 1
\end{array} \right\}
\left\{ \begin{array}{ccc}
1/2 & j_c & l_c \\
j_d & 1/2 & 1
\end{array} \right\} \\
\\
\times \delta_{l_a l_b}\delta_{l_c l_d}\delta_{J1}\delta_{T0} \;.
\end{array}
\end{equation}
We have calculated the pairing energy by taking the energy difference of states
calculated with the full Hamiltonian jj44b and the Hamiltonian $H_{eff}$
obtained by subtracting from jj44b Hamiltonian P01 or P10 pairing interaction. 
We have used $G$ =  0.276 for P01 and $G$ = 0.506 for P10 following refs. 
\cite{povesplb,zuker}. The contribution of  pairing energies  for $T=0$ odd spin
states and $T=1$ even spin states are shown in the figs. 4(a) and (b). Our results
are similar to the values obtained by Juodagalvis and \AA berg (JA)\cite{npa}.
However there are some important differences. Juodagalvis and \AA berg found that
the $T=1$, $J=0$ ground state has equal contribution of pairing energy from $T=0$ and
T=1 channels. However, we find that the two contributions are approximately
equal for $J \ge 4^+$ for both $T=0$ (odd $J$) and $T=1$ (even $J$) levels. For
the even $T=1$ levels with $J <4^+$, isovector pairing plays a larger role.
Similarly, for $T=0$ levels with $J < 5^+$, the isoscalar pairing plays a much 
greater role. JA also found a similar trend above the band crossing region.
However below the band crossing region, the isoscalar pairing is much larger
than the isovector pairing. They have obtained a measure of the contribution of
isovector pairing  mode. Due of isospin symmetry, the three parts of the
isovector pairing, $nn$, $pp$ and $np$, are identical in the $T=0$ states. Their
calculation shows that each component of the isovector pairing mode contributes
about 0.25 MeV to the $T=0,J=1$ state and in our case it is (see fig. 4a) 0.30
MeV.

From the studies of pairing energy for $T=0$ and $T=1$ bands,  Juodagalvis and \AA
berg have tried to understand how pairing energy plays an important role in
making the $T=1$ state to become the ground state. We have made a similar analysis
using our calculation. The isoscalar pairing  for the lowest $T=0$ and lowest $T=1$
bands is about 2 and 0.8 MeV. Similarly  the isovector pairing for the two cases
are 1 and 2.3 MeV. Thus the total pairing energy for $T=0$,$J=1$ is 3 MeV whereas
for $T=1$,$J=0$, the total pairing energy is 3.1 MeV. Thus our calculation also
shows that the $T=1$ band should be lower compared to the $T=0$ band because of the
gain of 0.1 MeV in pairing energy. It may be noted that our SM calculation with
jj44b interaction  predicts the $T=1$ and $T=0$ band head separation to be 148 keV
compared to the experimental value 571 keV whereas the MS interaction in JA
correctly  reproduces the separation. Hence compared to JA, our SM with jj44b
interaction shows a lesser favoring of pairing energy for $T=1$ ground state.

\section{Conclusions}

In the present work we have compared results of recently available experimental
data for $T=0$ and $T=1$ states for $^{62}$Ga within shell model and deformed
shell model results obtained using jj44b interaction. As discussed in detail in
Section 3, the SM and DSM explain the experimental data well. The analysis shows
that DSM with much smaller number of (deformed) configurations is adequate for
$^{62}$Ga. The present analysis goes much beyond the analysis presented before
for  $^{62}$Ga using SM and DSM in refs. [17,18] and adds more information. In
future, it is also important to improve further the effective interactions in
$f_{5/2}pg_{9/2}$ space and also include proton and neutron excitations across
the $Z=28$ shell by including the $1f_{7/2}$ orbital in the model space.\\

Further, the following broad conclusions can be drawn:\\

$\bullet$ The present shell model and the deformed shell model calculations 
with jj44b interaction describe the spectroscopic data reasonably well.

$\bullet$ We have calculated the $B(E2)$ values for isovector and isoscalar 
transitions. Comparison with one experimental data available for this nucleus
is good. The $B(E2)$'s for $T=1$ even spin states show a slow rising trend and 
finally decrease for $J=10^+$.

$\bullet$ Both SM and DSM with jj44b interaction showed a band crossing 
starting at $J=13^+$ for the lowest $T=0$ band (band crossing is the mechanism 
for back-bending). Past CSM calculations showed band crossing at $J=9^+$ 
\cite{npa}.

$\bullet$ We have calculated the pairing energy for $T=0$ and $T=1$ bands. We find
that for the lowest $T=0$ band, the contributions from isoscalar and isovector 
pairing are approximately equal for $J \ge 5^+$ while for $J <5^+$, isovector 
pairing plays a larger role. 

$\bullet$ The present shell model results correctly predict six 
$1^+$ levels with $T=0$ as reported in a recent experiment \cite{62gaprl}.

$\bullet$ We also obtain low level density at low energy for $^{62}$Ga in 
agreement with experiment.

$\bullet$  Future experimental investigations with radioactive ion beam
facilities will be required for generating more information about the structure
the $T=0$ and $T=1$ bands and the nature of $T=0$ and $T=1$ pairing
in heavier $N=Z$ odd-odd nuclei. \\

%\begin{acknowledgments}

The authors are thankful to Profs. B.A. Brown, R. Palit, David Jenkins and Dr. P.
Ruotsalainen for useful  discussions. RS is thankful to DST (Government of
India) for financial support.

%\end{fmffile}
\end{document}